\documentstyle[rotating,subfigure]{elsart}

\input{psfig.tex}

\begin{document}
\begin{frontmatter}

\title{	Characterization of Solidified Gas Thin Film Targets\\
via Alpha Particle Energy Loss}

\author[ubc]{M.C.~Fujiwara\thanksref{MCF}},
\author[uvic]{G.A.~Beer},
\author[tri]{J.L.~Beveridge},
\author[uvic]{J.L.~Douglas},
\author[gac]{T.M.~Huber},
\author[fri]{R.~Jacot-Guillarmod},
\author[sk]{S.K.~Kim},
\author[uvic]{P.E.~Knowles\thanksref{PEK}},
\author[wyo]{A.R.~Kunselman},
\author[uvic]{M.~Maier},
\author[tri]{G.M.~Marshall},
\author[uvic]{G.R.~Mason},
\author[tri]{F.~Mulhauser\thanksref{FM}},
\author[tri]{A.~Olin},
\author[psi]{C.~Petitjean},
\author[uvic]{T.A. Porcelli} and
\author[aus]{J.~Zmeskal}

\collab{Muonic Hydrogen Collaboration}

\address[ubc]{Department of Physics and Astronomy,
University of British Columbia, Vancouver, BC,
V6T 2A6 Canada}
\address[uvic]{Department of Physics and Astronomy,
University of Victoria, Victoria, BC, V8W 2Y2
Canada}
\address[tri]{TRIUMF, Vancouver, BC, V6T 2A3
Canada}
\address[gac]{Department of Physics, 
Gustavus Adolphus College, St. Peter, MN 56082, USA}
\address[fri]{Physics Institute,
Universit\'e de Fribourg, CH-1700 Fribourg, Switzerland}
\address[sk]{Department of Physics,
Jeonbuk National University,\\ Jeonju City 560-756, S.\ Korea}
\address[wyo]{Department of Physics,
University of Wyoming, Laramie, WY 82071, USA}
\address[psi]{Paul Scherrer Institute, CH-5232 Villigen, Switzerland}
\address[aus]{Austrian Academy of Sciences, A-1090 Wien, Austria}

\thanks[MCF]{Corresponding author, Tel: +1 604 222 1047 (ext. 6290), 
Fax: +1 604 222 1074, e-mail: fujiwara@triumf.ca}
\thanks[PEK]{Present address:
Universit\'e Catholique de Louvain, B--1348 Louvain--La--Neuve, 
Belgium}
\thanks[FM]{Present address: Physics Institute,
Universit\'e de Fribourg, CH-1700 Fribourg, Switzerland}

\begin{abstract}

A method is reported for measuring the thickness and uniformity of 
thin films of solidified gas targets.
The energy of $\alpha$ particles traversing
the film is measured and 
the energy loss is converted to thickness using the stopping power.
The uniformity is determined by measuring the thickness at different
positions with an array of sources. Monte Carlo simulations have been
performed to study the film deposition mechanism.
Thickness calibrations for a TRIUMF solid hydrogen target system are
presented.

\end{abstract}

\end{frontmatter}

\section{Introduction}

Recently, the use of thin films of solidified gases has 
attracted attention in particle beam experiments~[1-21].
A novel target system has been developed at TRIUMF for making 
films of solid
hydrogen isotopes and other gases for experiments in muon
catalyzed fusion~\cite{knowl93,knowl95,fujiw95b}.
Several measurements have been reported using the
apparatus~\cite{foste90,marsh93,jacot93,mulha94,mulha95,%
marsh95,knowl95b,jacot95,fujiw96} and 
more are proposed for the 
future~\cite{fujiw94,fujiw95,fujiw95c,jacot95b,bystr94}. 
Similar layers of solid hydrogen isotopes have been 
utilized in an attempt to
produce a source of low energy negative 
muons~\cite{nagam89,stras95,stras96}.  
Production of a muonic helium ion beam 
of keV energies~\cite{nagam93b}
as well as MeV energies~\cite{fujiw94,fujiw95} 
has been proposed
using multilayer cryogenic films.

In these experiments, characterization of the target films is often 
important in the analysis of the experimental data. For cross section
measurements, in particular, the accurate knowledge of the thickness and
uniformity of the target films is essential, since the uncertainty in
the thickness directly propagates to the final results.  

With the TRIUMF target system, 
solid layers of hydrogen isotopes (protium $^{1}$H$_{2}$, 
deuterium, tritium and their mixtures)
and other gases such as neon can be made with thicknesses
ranging from a few $\mu$g$\cdot$cm$^{-2}$ ($\sim$1 $\mu$m for protium)
to a few mg$\cdot$cm$^{-2}$.
Due to the spatial and cryogenic limitations, 
conventional methods for thin film
thickness measurement, such as optical interferometry, cannot be used
with our system.

Thickness measurements of condensed gases have been reported by several
authors.
S\o rensen {\it et al}.\ used a quartz crystal oscillator to 
measure solid
hydrogen film thicknesses~\cite{soren76}. However, 
the method suffered from 
a severe non-linearity, 
which was attributed to the low density of hydrogen.
Rutherford Backscattering (RBS)~\cite{chu78b} which was used, for
example, by Chu {\it et~al.}~\cite{chu78} 
to measure the thickness of solid argon,
oxygen, and CO$_{2}$, is kinematically impossible for protium targets.

In the present work,
we have used the energy loss of $\alpha$
particles traversing the film to measure the thickness.
Uniformity was determined by measuring the thickness at different
positions with an array of sources.
Following the introduction, details of the experimental
apparatus are described 
in section 2. Sections 3 discusses 
data analysis and uncertainties.
The results of various measurements are presented in section 4, and
comparison with Monte Carlo simulations, as well as discussion is
given in section 5.

\section{Experimental apparatus}

\begin{figure}
\begin{center}
\mbox{\psfig{figure=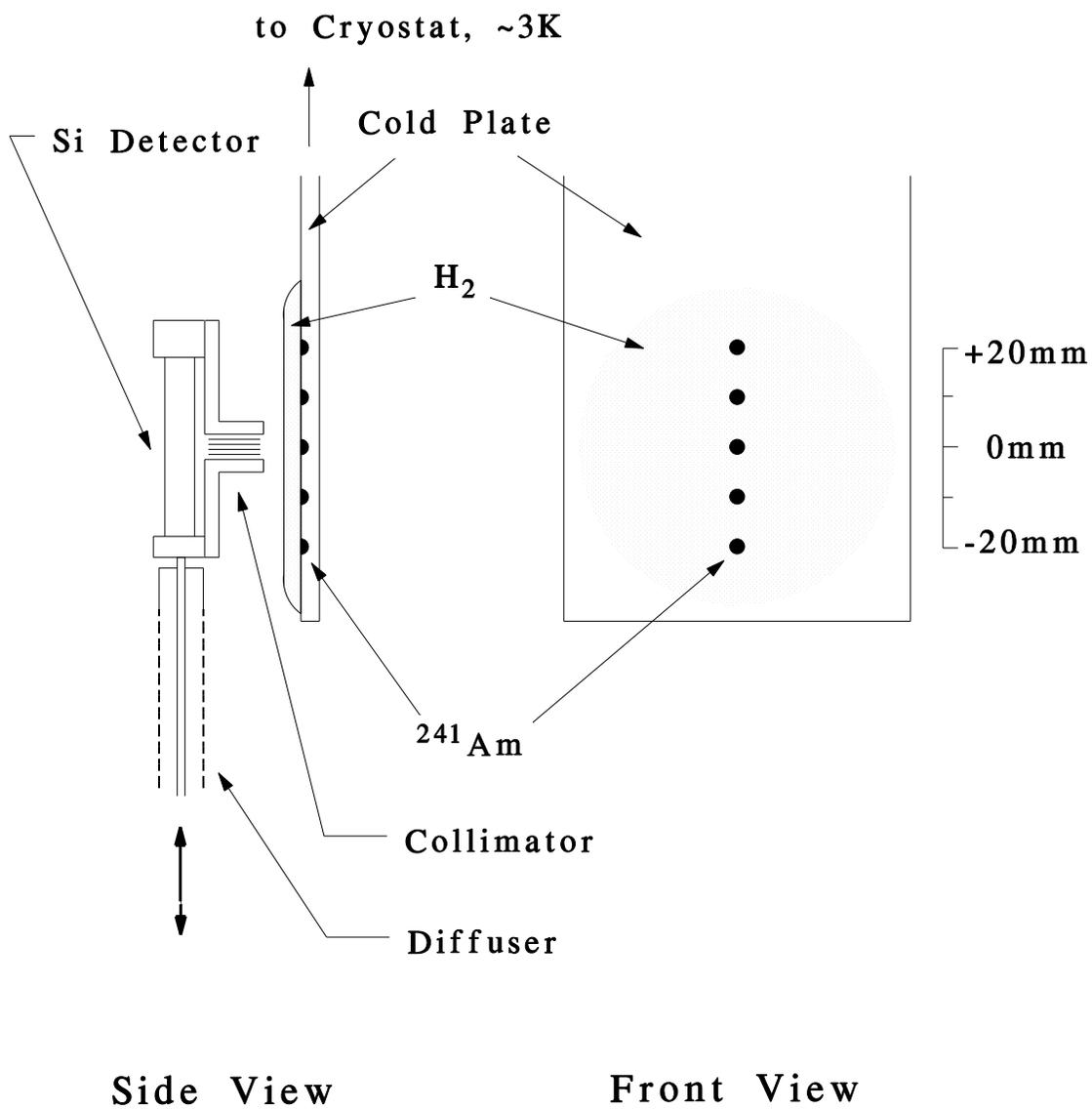,%
height=15cm,clip=}}
\end{center}
\caption{Schematic view of the experimental setup with solid hydrogen
(H$_{2}$) layer. The apparatus is contained in an evacuated chamber.}
\label{fig:set-up}
\end{figure}

Figure~\ref{fig:set-up} shows a schematic view of the experimental
apparatus. 
The general description of the target system including gas deposition
mechanism is given in Refs.~\cite{knowl93,knowl95}. 
A linear array of five alpha spot source was 
custom-manufactured\footnote{Isotope Products
Laboratories, 1800 N.\ Keystone St., Burbank, CA, USA.} by
electro-deposition of $^{241}$Am onto a gold-plated oxygen-free copper
plate. The spot sources had nominal diameters of 3 mm, 
a center-to-center
separation of 10 mm, and were covered by a thin gold layer
($\sim$200 $\mu$g$\cdot$cm$^{-2}$) for safety and ease of handling.
The plate, enclosed in an evacuated chamber,
was cooled to approximately 3~K, and hydrogen
gas solidified onto it when introduced through the gas diffusing 
mechanism (diffuser).
Different gas deposition systems have been developed to 
incorporate experimental demands, and to 
improve the performance~\cite{knowl93,knowl95}.
For system 1, the gas was released
through a diffuser made of a thin
stainless steel foil perforated with many holes ($\sim$0.2 mm
diameter). System 2 used the same diffuser foil, but as a result of
gas line modifications to incorporate tritium compatibility, it 
had gas inlet tubing with a higher volume, which acted as an 
unwanted buffer
volume as we shall see in section~\ref{sec:linearity}.
System 3 is the latest version with 
the gas inlet lines replaced with low volume, high
conductance tubing, and a sintered metal diffuser (2
$\mu$m porosity) employed to ensure microscopic homogeneity of
deposition.

The distance between the diffuser surface
and the target plate, which is adjustable depending on the experimental
requirements, 
was for the present measurements
about 14 mm for System 1 and 2, and about 8 mm for System 3.

The amount of gas injected into the system was measured in units of
Torr$\cdot$litre (abbreviated T$\cdot l$), where one T$\cdot l$ 
corresponds to the 
number of molecules in one litre of gas at a pressure of 1 
Torr and ambient
temperature ($\sim$295 K). This unit was operationally 
convenient since 
the number of molecules can be compared, independent of the isotopic 
composition.

The rate of gas deposition was typically of the order of a few 
(T$\cdot l)\cdot s^{-1}$ or less, the limit imposed by a requirement 
to avoid significant
intermolecular interactions which would lead to heat conduction from
the relatively hot ($\sim$100 K) diffuser to the cold plate ($\sim$3
K).

\begin{figure}
\begin{center}
\begin{sideways}
\mbox{\psfig{figure=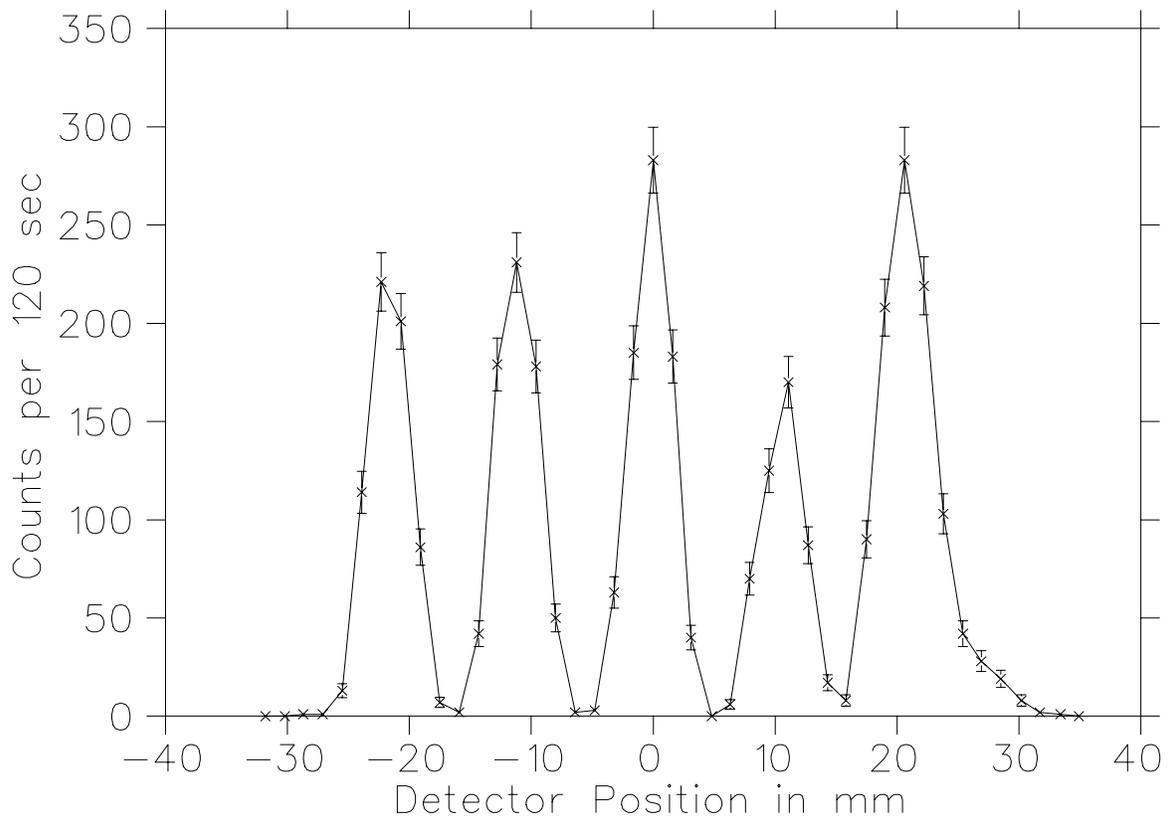,height=6in}}
\end{sideways}
\end{center}
\caption{Counts versus vertical position of the silicon detector
	with respect to the center. %
	Each peak corresponds to one of the five source spots. %
	The detector is collimated such that it sees only one 
	spot at a time.}
\label{fig:pos-counts1}
\end{figure}

Alpha particles penetrating the hydrogen film were detected 
by a passivated, implanted planar silicon 
detector\footnote{Canberra, model FD/S-600-29-150-RM.}
of active thickness
150 $\mu$m and area 600 mm$^{2}$.
The detector was mounted on the diffuser frame which 
was part of a mechanism
that allowed the diffuser to be inserted and retracted
(Fig.~\ref{fig:set-up}). The detector thus
moved vertically to allow a measurement of the thickness at
different positions by detecting the $\alpha$ particles from each of
the five spot sources. A collimating device which consisted of an array
of small holes (diameter $\sim$1 mm) restricted the
angular path of the $\alpha$ particles to accept alphas from only  
one spot source at a time. 
The signal from the detector was recorded with a standard spectroscopy
system consisting of a charge sensitive preamplifier, 
linear amplifier, and
analog-to-digital converter, together with a CAMAC/VAXstation data
acquisition system.
The energy scale of the alpha detection system was calibrated using a
separate $^{241}$Am source. Calibration was frequently required
since temperature
variations in the detector could cause significant shifts in the gain. 
The system achieved a
typical resolution of 0.4\% (FWHM) at a detector
temperature near 100 K.
The profile of alpha counts 
versus the vertical position of the collimated detector 
for a bare target 
(i.e.\ no solid hydrogen layer) 
is shown in Fig.\ \ref{fig:pos-counts1}. The plot confirms that we
detected $\alpha$ particles from only one source spot at a time.

\section{Analysis}

\subsection{Thickness determination}

\begin{figure}
\begin{center}
\begin{sideways}
\mbox{\psfig{figure=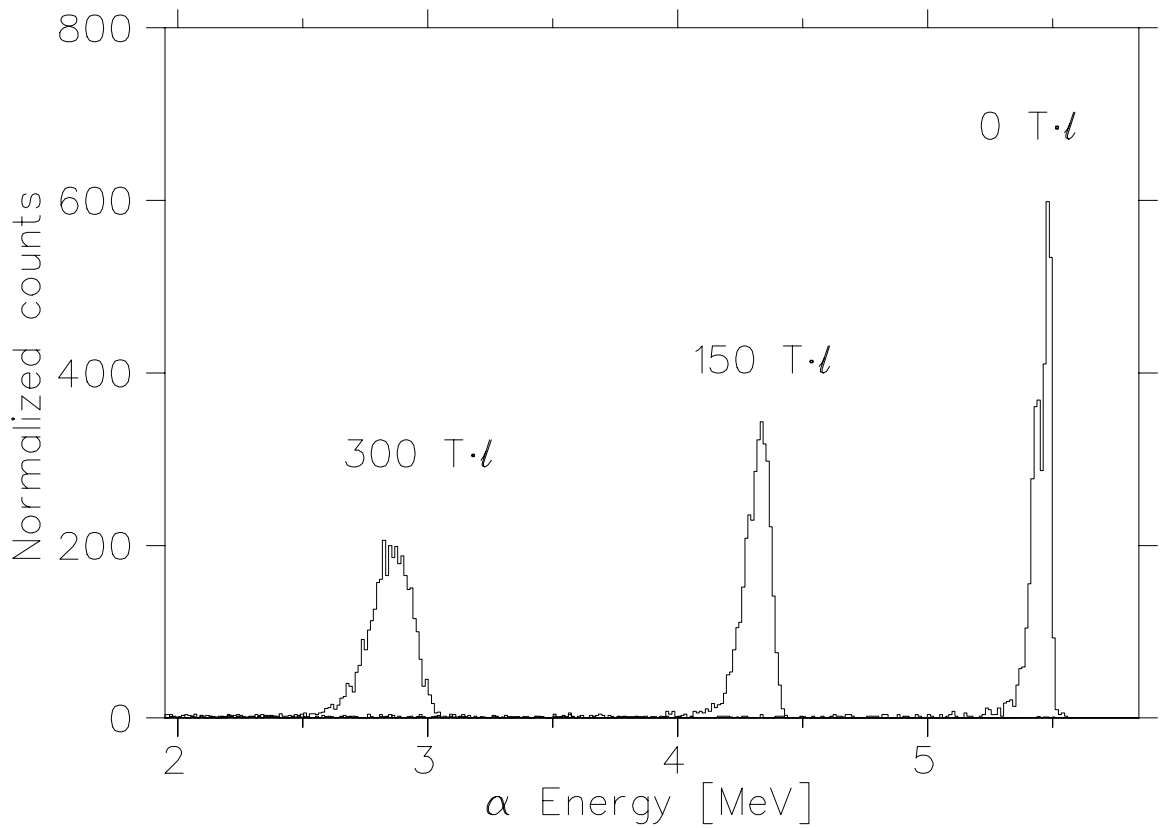,height=6in}}
\end{sideways}
\end{center}
\caption{Alpha particle energy spectra with different 
	thicknesses of hydrogen 
	film, where the peaks are normalized to the same number of
	counts.
 The numbers above each peak indicate the amount of hydrogen 
 gas injected.}
\label{fig:spectra}
\end{figure}

Figure~\ref{fig:spectra} shows an example of the energy spectra of
$\alpha$ particles penetrating hydrogen films
with different amounts of gas injected, namely 0,
150, and 300 T$\cdot l$. 
The shift of the peaks 
to lower energy with increased injected gas is clearly
visible.
Note also 
the peak broadening (due mainly to straggling) and the 
asymmetric peak shape 
which is due in part to the energy loss in 
the protective gold layer on the source.
We determine the mean energy 
value $<\!E\!>$ from the centroid of the
energy distribution $f(E)$ in the spectrum 
via
\begin{equation}
<\!E\!> = 
\frac{{\displaystyle\int_{<\!E\!>-\epsilon}^{<\!E\!>+\epsilon}} 
f(E) E dE}
{{\displaystyle \int_{<\!E\!>-\epsilon}^{<\!E\!>+\epsilon}} f(E)dE},
\label{eq:centroid}
\end{equation}
where $\epsilon$ is a finite cutoff value. 
The use of the centroid in the analysis achieves sufficient accuracy
while
avoiding the difficulties in  
fitting the irregular
peak shapes, which vary depending on the source spot and target
thickness.

In the approximation that the angular dispersion of $\alpha$ 
particles is
avoided due to the use of a collimator,
the thickness of the target $T$ can be obtained from 
$R(<\!E\!>)$,
the alpha 
range as a function of energy:
\begin{equation}
T = R(<\!E_{init}\!>) - R(<\!E_{fin}\!>),
\label{range}
\end{equation}
where $<\!E_{init}\!>$ is the initial energy of the $\alpha$ 
particles and
$<\!E_{fin}\!>$, the
energy after traversing the target.

For our analysis, the stopping power or the range in the  
{\it solid state} of hydrogen for $\alpha$ particles in the 
energy range
of $\sim$2--5 MeV was needed, but no experimental data is available
for solid hydrogen at these energies.
The detailed discussion of 
the effect of physical phase on stopping
power for heavy charged particles, which is reported for keV
projectiles in hydrogen~\cite{borge82} and in nitrogen~\cite{borge82b},
as well as for MeV ions in organic and other 
materials~\cite{ziegl76,thwai78,thwai83}, 
can be found in Ref.~\cite{fujiw94} (see also
reviews~\cite{borge85,thwai85,thwai87}).
After a critical survey of the literature~\cite{fujiw94}, 
a recent compilation for gaseous hydrogen 
by the International Commission of Radiation Units
and Measurements (ICRU)~\cite{icru93} 
was used in the analysis.

\subsection{Uncertainties}
\label{sec:uncertainties}

The systematic uncertainties considered in the analysis include 
knowledge of stopping power, the effect of energy cuts, and energy 
calibration of the detector. The uncertainty in the stopping power is
difficult to estimate, since neither experimental data nor
satisfactory theory is available for our case. The ICRU
table~\cite{icru93} claims
an accuracy of $\sim$1-2\% at 4 MeV and  $\sim$2-5\% at 1 MeV
for gaseous hydrogen. 

Ziegler et al.,
in another commonly used compilation~\cite{ziegl85}, 
give an estimate of stopping power of solids 
for which there are no data available, by 
interpolating (or, for hydrogen, extrapolating) 
the data from other elements
under certain assumptions.
They
quote a 5\%
average accuracy for $\alpha$ particle stopping powers in solids, 
which is simply
the average 
of deviations taken from the collections of experimental data in
the literature for many elements.

The two tables (\cite{icru93} and
\cite{ziegl85}) agree with each other 
within 3\% at $\sim$ 2 MeV,
and the thicknesses derived using both tables 
agree within 2\% (the differences in stopping powers partly cancel 
with one
another upon integration over energy).
We shall conservatively 
quote 5\% of the derived thickness as the uncertainty due to 
the stopping
power, including a possible physical phase effect.
This was the limit on 
our accuracy in most cases.

The effect of the finite cutoff values ($\epsilon$ in
Eq.~\ref{eq:centroid})
in determining the centroid of
the energy spectra 
was investigated by changing the cuts
and its uncertainty was found to be small compared to that from
the stopping powers in most cases. 
The uncertainty in the detector energy calibration is estimated to be 
less than 2 keV,
except for a few
measurements during which a temperature variation in the silicon
detector resulted in a
gain shift.

These systematic uncertainties, as well as the statistical
uncertainties,
were added quadratically to obtain the total uncertainty. 
Except for very thin layers, 
whose thicknesses have to be determined from a
small difference in the initial and final alpha energies, the
knowledge of the stopping power dominated the uncertainty.

\section{Results}
                     
\subsection{Linearity of film deposition}
\label{sec:linearity}

\begin{figure}
\begin{center}
\begin{sideways}
\mbox{\psfig{figure=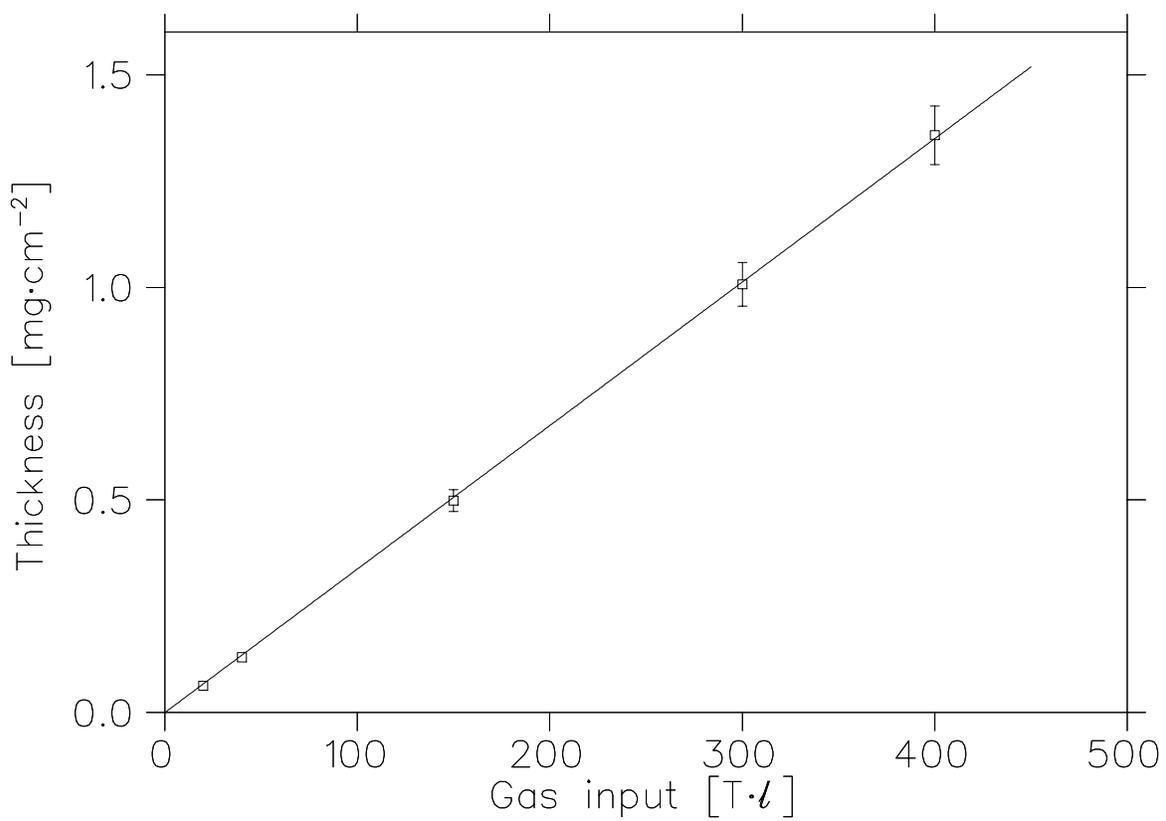,height=6in}}
\end{sideways}
\end{center}
\caption{Test of the linearity of deposition. 
The solid line represents
a least-squares fit
to the data points plotted with error bars.}
\label{fig:linearity}
\end{figure}

The linearity of the deposition was tested by measuring the thickness of
films made with different amounts of gas input under different
conditions. The data shown in Fig.~\ref{fig:linearity} were taken at
the central source position (0 mm) for films made by diffuser
system 1.  Some of the films  (40, 300 T$\cdot l$) were made by
depositing gas on top of existing films, whereas others  (20, 150,
400 T$\cdot l$) were made with a single deposition. 
The solid line
represents a weighted
least-squares fit to the data (plotted with error 
bars\footnote{Error bars presented in this paper represent 
total uncertainties including that 
from stopping power, unless otherwise stated.}).
The slope represents the conversion factor
from gas input to the film thickness at this source position (0 mm). 
A possible non-linear 
component in deposition was
investigated by allowing a quadratic term in the fitting function. 
The maximum allowable non-linear contribution, when extrapolated to a
1000 T$\cdot l$ target, was similar in size to the uncertainty in the
thickness due to the stopping power.
Similarly, the data from other source positions on the plate, and with 
diffuser systems 2 and 3, showed good linearity for these moderate 
thicknesses.

The extrapolation of the results to very thin films requires some
caution; if a small amount of the gas remained in the gas transfer tubes,
it would be lost from the layer, giving a small offset 
in gas deposition. 
This gas loss is negligible for
thick layers, but can be important for thin layers.
The effect was examined for systems 2 and 3 by 
comparing two series of measurements; 
(1) thick films made by a
single deposition of a large
amount of input gas, in which the gas loss is negligible, and
(2) thin films made by sequential 
depositions of small amounts of gas, where
the gas loss from each deposition, if it exists, 
is multiplied to give a measurable effect after several 
depositions.

Due to the small energy loss,
measurement of very thin films was difficult, and we have assumed that
the same cutoff value, $\epsilon$ in Eq.~\ref{eq:centroid}, 
can be used to determine both the initial and the
final energies (this is not generally valid in the thick film
measurement due to the peak broadening). 
Provided that the same value is used for both initial and final
spectra, the choice of $\epsilon$ did not affect the resulting thickness
values.

\begin{figure}
\begin{center}

\subfigure[System 2]{%
\begin{sideways}
\mbox{\psfig{figure=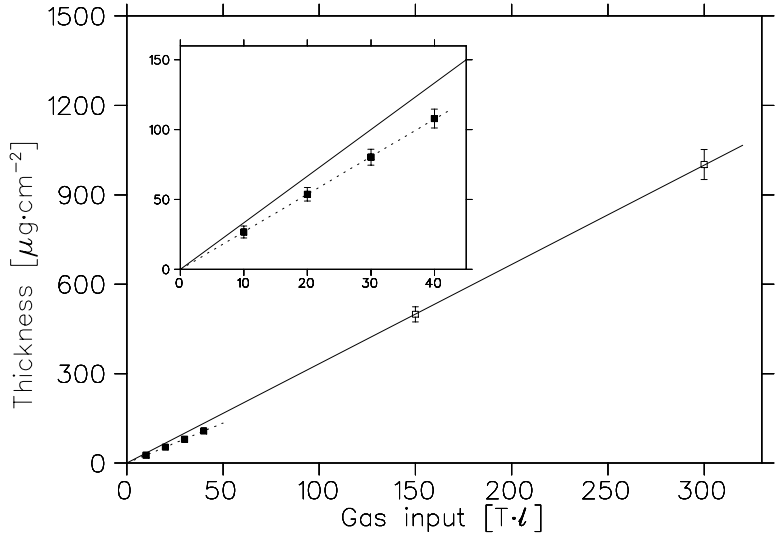,%
height=4in}}
\label{fig:thin_r3}
\end{sideways}
}
\subfigure[System 3]{%
\begin{sideways}
\mbox{\psfig{figure=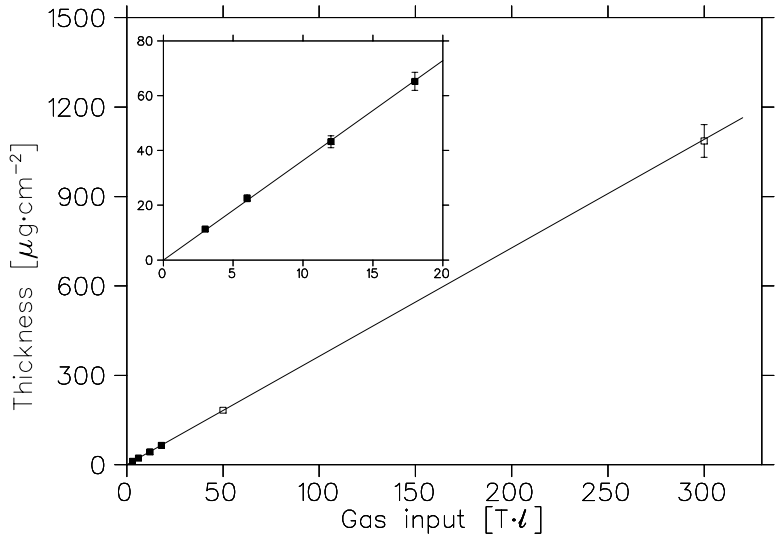,%
height=4in}}
\label{fig:thin_r4}
\end{sideways}
}
\end{center}
\caption{Sequential deposition of very thin films shown as filled 
squares is 
compared to the thick 
film deposition shown as open squares for different systems:
(a) system 2, where the difference in slopes between the two series of
depositions indicates the gas loss effect, and 
(b) system 3, showing no evidence for such an effect.
Inserted boxes in the figures illustrate details at small thicknesses.}
\end{figure}

Figures \ref{fig:thin_r3} and \ref{fig:thin_r4} show the results of the
comparisons for systems 2 and 3 respectively.
For system 2 (Fig.~\ref{fig:thin_r3}), 
sequential deposition of small amounts of gas (measurement (2), filled
squares) 
resulted in a smaller
thickness per unit gas input than thick films made with a single
deposition (measurement (1), open squares), 
indicating that when making very thin films a non-negligible
amount of the gas remains in the gas transfer tubes without being 
deposited. On the other hand,
System 3 (Fig.~\ref{fig:thin_r4}), which was in fact designed to 
remove the effect,
shows no evidence of such gas loss.

\subsection{Thickness profile}
\label{sec:profile}

\begin{figure}
\begin{center}
\begin{sideways}
\mbox{\psfig{figure=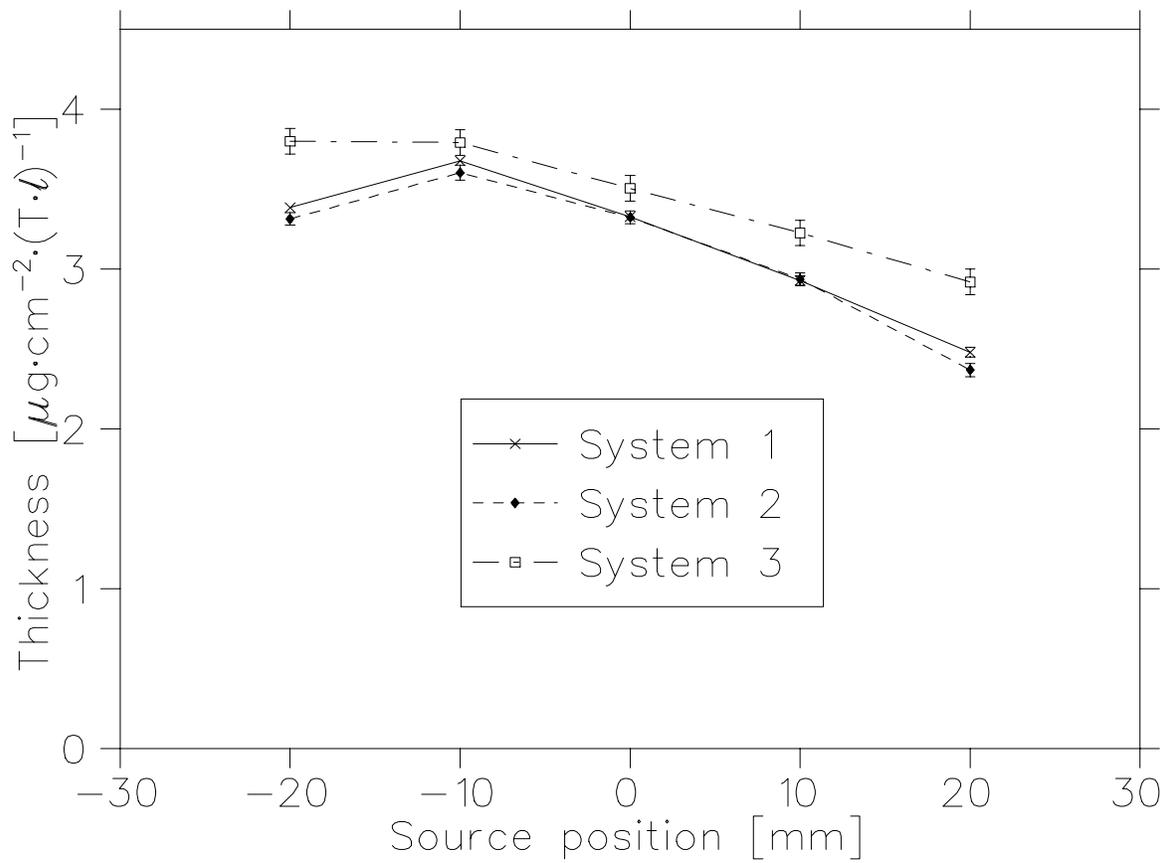,height=6in}}
\end{sideways}
\end{center}
\caption{Thickness profiles for different diffuser systems. 
Thickness per unit
gas input is plotted against the source positions. 
Because these are relative measurements, 
the error bars do
not include the uncertainty from the stopping power.}
\label{fig:profile}
\end{figure}

Shown in Fig.~\ref{fig:profile} are 
the thickness profiles from different deposition systems.
An asymmetric non-uniformity with respect to
the center can be observed. The error bars do not include the
uncertainty from the stopping power, since it is not relevant when 
considering relative uniformity. 

One may assume that the thickness of a film at a particular position
depends only on the relative distance from the diffuser,
which is the case if, for example, the molecules stick 
onto the cold plate at the first contact
as suggested from the measurements described in
Section~\ref{sec:other}.
In this model, the film profile should have translational invariance
under the diffuser displacement.

\begin{figure}
\begin{center}
\begin{sideways}
\mbox{\psfig{figure=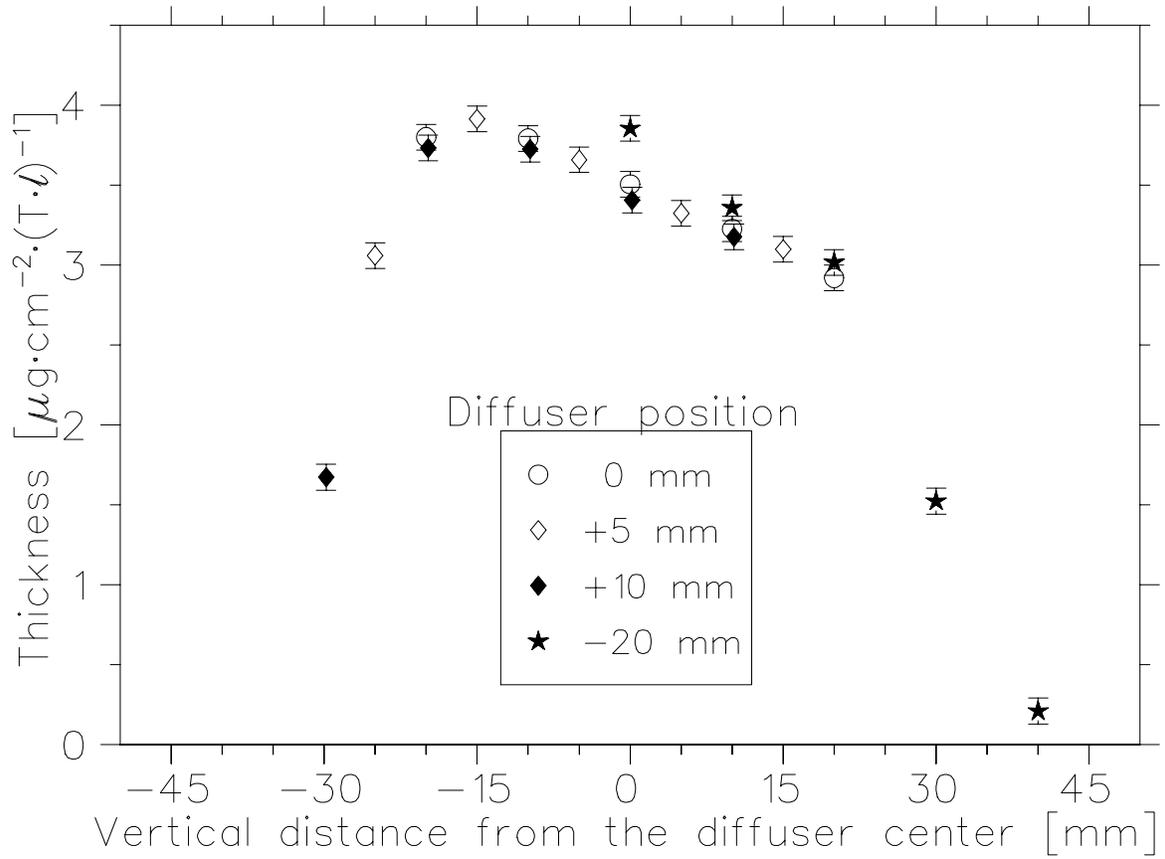,height=6in}}
\end{sideways}
\end{center}
\caption{Film thicknesses for system 3 are plotted 
against the vertical distance from the
diffuser center (diffuser coordinates).
The error bars do
not include the uncertainty from the stopping power. See text for the
detail.}
\label{fig:relative}
\end{figure}

A series of measurements were made using system 3 with films that
were deposited with the diffuser displaced by 0 mm, +5 mm,
+10 mm and -20 mm from the nominal standard position, the positive
direction being upward. In Fig.~\ref{fig:relative}, the resulting
thicknesses are plotted against {\em diffuser coordinates}, i.e.,
the relative vertical distance from the center of the diffuser, unlike 
Fig.~\ref{fig:profile} which was plotted against the distance from the
center of the source plate.
The data from different films, thus plotted, are consistent,
indicating the validity of translation invariance of the film profile.
One exception is the filled star point 
at distance 0 mm in the diffuser coordinates, 
which represents the thickness at the center of the diffuser 
for a film deposited with the diffuser retracted 20 mm downwards.
At this position the diffuser, which has an 
active diameter of 60 mm, was aligned with the  
bottom source spot (see Fig.~\ref{fig:set-up}) and a significant
fraction of emitted gas molecules missed the cold plate, hence 
the molecules
may bounce around inside the vacuum system, eventually 
sticking nearer to the bottom rather than the top of the cold plate.
Nevertheless, validity of the translational invariance indicated above 
suggests that it is possible to measure a thickness at an 
arbitrary point
along the vertical axis with the present technique,
despite the fact that $\alpha$ particles are emitted only from
discrete source spots.
Figure~\ref{fig:relative} can thus be considered as a good
representation of the thickness profile of the film made with system 3.
Based on these results,
the calibration factors (thickness per unit gas input) for system 3
at each position with respect to the diffuser center, including the 5\%
uncertainty from the stopping power,
are given in Table~\ref{tab:conversion}.
When more than one measurement exist for the same position, an 
average was taken, except that
some points from the diffuser position of -20 mm, which may
be in error as discussed, 
were not included in the final
average.

It should be noted that, in general, the
calibration factor depends on the distance between the diffuser surface
and the cold plate surface which, in the present case, was 8 mm.
For system~1, which had a 14 mm separation between the diffuser and
cold plate,
the calibration factors
at the standard vertical diffuser
position are given in Table~\ref{tab:conversion_r1}.
The values for
system 2 are similar and agree with system 1 within the
uncertainty (see Fig.~\ref{fig:profile}).

Thicknesses of deuterium and tritium in units of 
$\mu$g$\cdot$cm$^{-2}$ are 
factors of 2 and 3, respectively, larger than for a protium film 
with the same number of molecules due to the isotopic mass difference. 
The measurement of a deuterium film, when corrected by this factor, 
showed good agreement with protium films.

\subsection{Other measurements}
\label{sec:other}

Measurements were made with films deposited under different conditions
to see the effects on thickness and uniformity.
No deviation was found for films made with and without pumping
the target vacuum during deposition within the relative uncertainty of
about 1\%.
Reducing the deposition rate by an order of magnitude 
also did not noticeably affect either thickness or uniformity. 

The gas deposition system is capable of making a second film on a
separate cold surface 
through the opposite side of the diffuser~\cite{knowl95}.
The apparatus was designed to minimize unwanted 
cross deposition from one side to
the other by shielding with cold surfaces.
By intentionally releasing a large amount of gas through the opposite
side, cross deposition on the spot source target was checked.
The measurement, made with similar assumptions on the cutoff value
$\epsilon$ to the measurement of very thin films described in 
Section~\ref{sec:linearity}, 
indicated that less than one part in a thousand of the injected 
gas arrived at the central spot.
This result, together with the fact that pumping the system did 
not affect
deposition, suggests that the gas molecules are likely to 
stick to the cold surface on first contact, which would explain the
observed translational invariance of film profiles
(Section~\ref{sec:profile}).

Beam experiments using solid hydrogen targets sometimes last for a few
days, so it is important to check the effect of target aging.
The results of measurements, made 8 hours apart, of the same film   
were consistent with each other, giving an upper limit in thickness
variation $\Delta T/T  \leq 0.5 \%$ over 8 hours. In the analysis, 
a similar assumption 
in the cutoff value was made.

Films of other gases can also be deposited with the target system (for
example, neon films have been used in TRIUMF
experiments~\cite{jacot93,jacot95}).
Thickness measurements of neon films were made using two different gas
inputs at three of the five source positions.
The stopping power in gaseous neon taken from the ICRU 
tables~\cite{icru93} was used to convert
energy loss to thickness.
The same 5\% uncertainty in stopping power was assumed.
Calibration factors for the three spots are given in
Table~\ref{tab:neon}.

\begin{figure}
\begin{center}
\begin{sideways}
\mbox{\psfig{figure=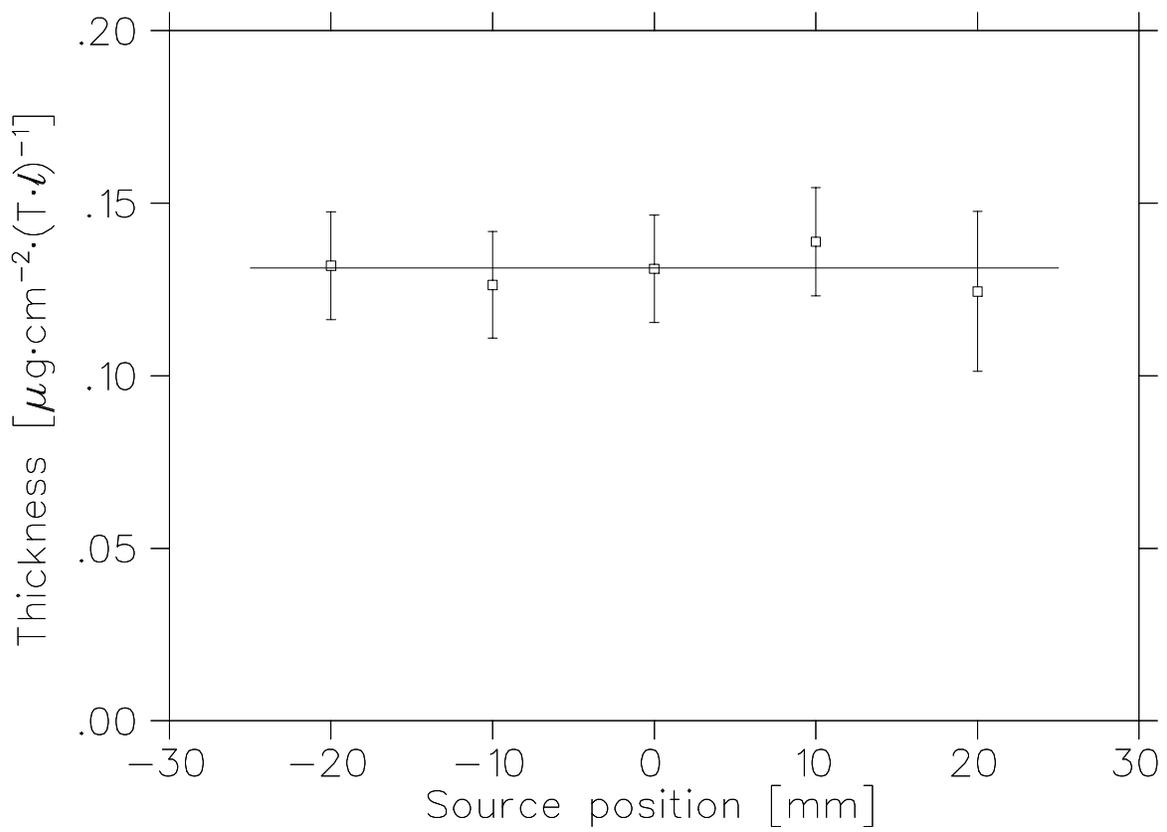,height=6in}}
\end{sideways}
\end{center}
\caption{Alternative deposition: Hydrogen gas was 
introduced to the entire
cryostat with
vacuum pump closed. The gas stuck uniformly 
to all of the cold surfaces.}
\label{fig:alternative}
\end{figure}

Another method of deposition was tested using system 1; 
the diffuser was completely moved away from the cold target plate, and a
relatively large amount of gas (300 T$\cdot l$)
was introduced into the system with all the pump valves closed.
The vacuum pressure during deposition was kept relatively 
high to allow some
interactions among the molecules.
The thickness profile of the film is 
shown in Fig.\ \ref{fig:alternative}.
Notice the much lower deposition efficiency compared to 
normal film deposition shown in Fig.\ \ref{fig:profile}. 
The uncertainties are relatively large due to the small thickness and a
noise problem experienced during some of the measurements.
Nevertheless, the results suggest a more uniform film deposition,
so this approach may be useful in
certain cases in which uniformity is critical.

\section{Discussion}
\label{sec:discussion}

\subsection{Monte Carlo simulation}

\begin{figure}
\begin{center}
\begin{sideways}
\mbox{\psfig{figure=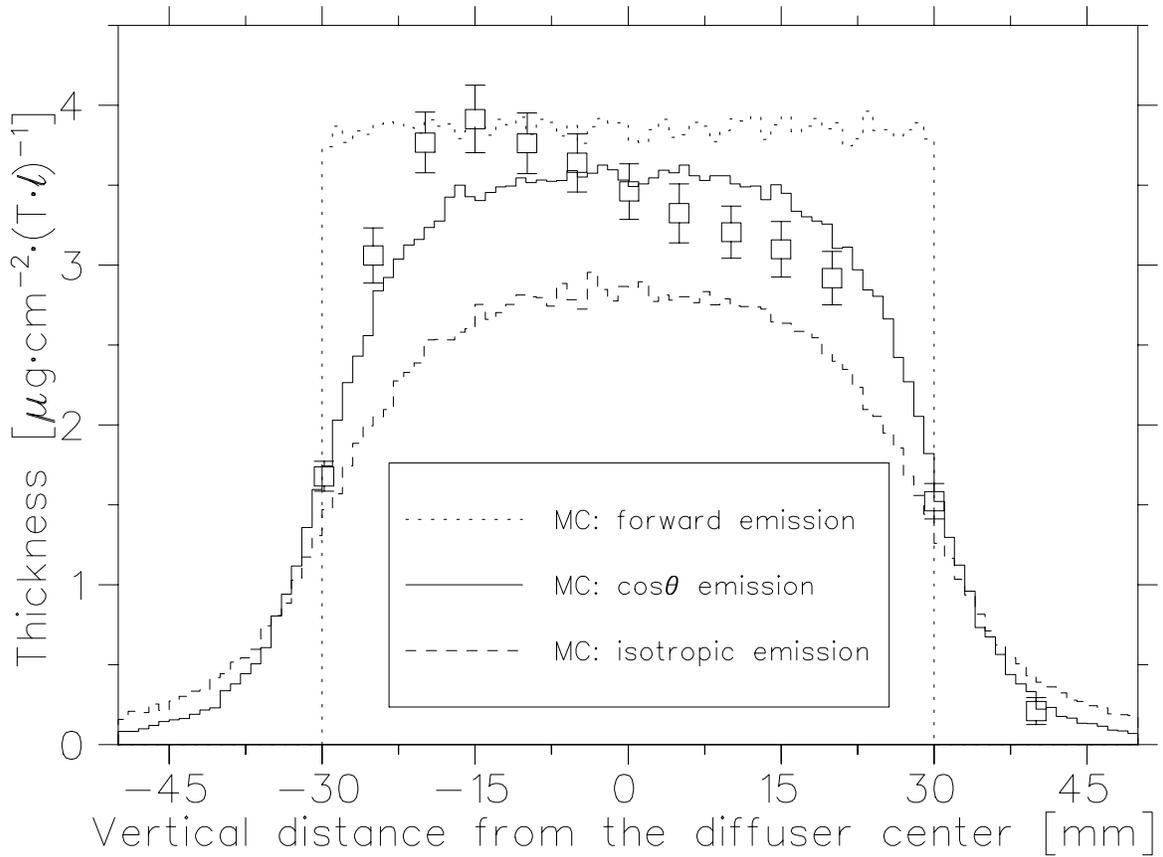,height=6in}}
\end{sideways}
\end{center}
\caption{The data, with error bars
which include the uncertainty from the stopping power, 
denote the averaged thickness profile for system 3 plotted against the
relative vertical distance from the diffuser center.
The histograms are the the simulated thickness profiles from Monte Carlo
calculations with different assumptions for the angular distribution of
molecules emitted from the diffuser surface.}
\label{fig:mc}
\end{figure}

In order to better understand the mechanism of gas deposition,
Monte Carlo simulations were performed 
with the following assumptions:
(1) molecules are emitted
uniformly from the gas diffuser surface,
(2) the molecules stick to the cold surface at the position of first
contact, and 
(3) there is no interaction between the molecules.
The last assumption is justified since the requirement 
to keep the target film temperature cold
demands an insignificant intermolecular interaction, minimizing  
heat conduction from the warmer diffuser to the target plate. 
Pressure must be maintained low enough during deposition 
to keep the 
mean free path of the emitted molecules comparable to
or larger than
the diffuser--to--plate separation.
Three different models for 
the angular distribution of molecules emitted from the diffuser surface
were used in simulations;
(a) the (unrealistic) forward emission model assuming $\theta = 0$, where
$\theta$ is the angle with respect to the normal 
to the diffuser
surface, (b) the isotropic emission model 
assuming that the number of molecules emitted into unit solid angle $dw$
is constant, independent of $\theta$, i.e.\  
$dw/d(\cos \theta) =$ const.,
and (c) the $\cos \theta$ emission model using a diffusion-like angular
distribution 
$dw /d(\cos \theta) \propto
 \cos \theta$.
The diffuser diameter of 60 mm and the distance of 8 mm 
between the diffuser and cold plate surface  
were used in simulations to compare with the measurements for 
system 3.  
There are no free parameters in the simulations
other than the emission angular distribution.
Figure~\ref{fig:mc} compares the simulation results with 
the film thickness profile for system 3 as deduced above, 
which is plotted
with error bars that now include the uncertainty from stopping power.
The agreement with the $\cos \theta$ model (c) is quite good,
except for the asymmetry in the shape. 

The asymmetry, which obviously cannot be reproduced with our 
simulations, 
is unlikely to be due to an effect of gravity on the target, 
since the relative non-uniformity remains constant over a wide 
range of the thickness.
The shape may be partly explained by the fact
that the gas is introduced from the bottom of the diffuser
system, hence gas molecules may have a larger probability of diffusing
out at the bottom rather than the top of the diffuser, indicating a
breakdown of assumption (1). 

Similar simulations with the diffuser--to--plate distance taken to be 
14 mm were compared to
data for system 1. The $\cos \theta$ model
best describes the data,
but comparison
indicates a slightly more forward peaked emission angular distribution.
It should be recalled, however, that system 1 had a different 
perforation structure from system 3.

\begin{figure}
\begin{center}
\begin{sideways}
\mbox{\psfig{figure=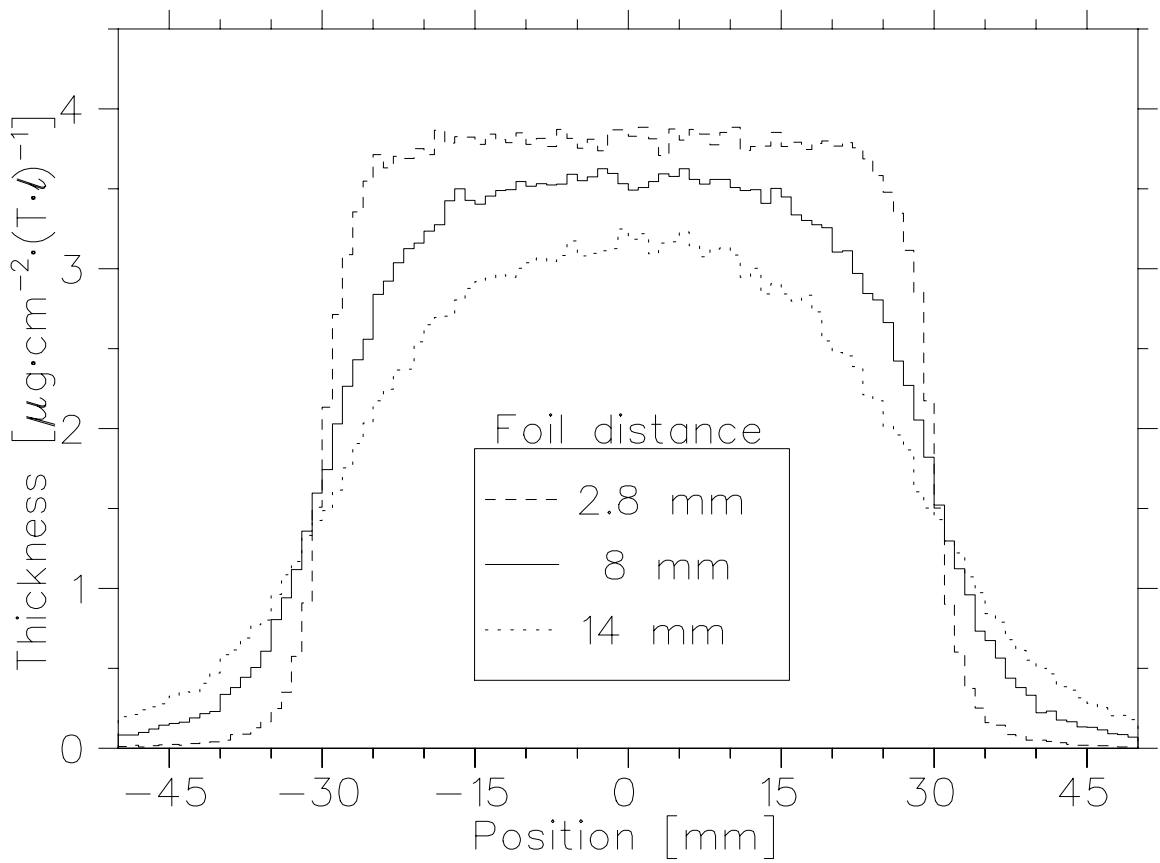,height=6in}}
\end{sideways}
\end{center}
\caption%
{Monte Carlo simulations comparing different distances between 
the diffuser
surface and the cold foil surface.}
\label{fig:mc_distance}
\end{figure}

Reasonable success in the simulation with the $\cos \theta$ model 
gives us some confidence in scaling the
thickness as a function of distance between the diffuser surface and the
target support foil surface. 
Figure \ref{fig:mc_distance} shows a comparison of the 
simulated thickness
profiles with different distances.

\subsection{Effective thickness}

The results given in Tables~1--3
should be used with caution
when the films are used as a target for beam experiments. 
Since the present work measures the profile only in the vertical
dimension, the profile in the horizontal dimension has to be assumed
when estimating an average target thickness.
For a non-uniform layer, 
the average thickness depends on the width and profile
of the beam which stops in the target.
The angular divergence of the
beam also contributes to the effective thickness.

An effective average thickness can be defined via
\begin{equation}
	T^{eff} = \frac{\sum w_{i} T_{i}} {\sum w_{i}},
\end{equation}
where $T_{i}$ is the thickness at the $i$th measured spot, and 
$w_{i}$ the weighting factor.
A weighted root--mean--square deviation of thickness
is defined via:
\begin{equation}
	\Delta T^{eff}_{rms} = \sqrt{ \frac{\sum w_{i} 
			(T_{i}-T^{eff})^{2}} 
			{\sum w_{i}}}.
\end{equation}
This is a quantitative measure of the 
non-uniformity and is useful when optimizing 
the vertical position of the diffuser for deposition.

The weighting factors depend on the beam parameters and the type
of measurement.
For example, when the target is used to stop particle beams from 
an accelerator as in Ref.\ \cite{foste90}, 
a gaussian distribution with a certain width may be justified.
However, when muonic hydrogen atoms emitted from a solid hydrogen layer
are stopped in a deuterium film, as in Ref.\ \cite{mulha95}, 
the atomic beam is divergent with an angular distribution 
close to $dw /d(\cos \theta) \propto \cos \theta$, and
depends on scattering cross sections.
In this case the detailed averaging has to be done with 
a Monte Carlo
simulation which includes differential scattering cross sections. 

\section{Conclusions}

Deposited films of solid hydrogen isotopes and neon 
have been characterized
via the energy loss of $\alpha$ particles. 
The present method can be applied
to a relatively wide range of film thicknesses, {\it e.g.} for protium,
from $\sim$1 $\mu$g$\cdot$cm$^{-2}$ to $\sim$1 mg$\cdot$cm$^{-2}$.
The accuracy of the measurement
is limited by the uncertainty in the stopping powers, but the relative
accuracy can reach better than 1\%.
The uniformity can be measured by sampling the thickness at 
different positions with an array of alpha sources. Furthermore, 
it was possible to determine the thickness at an arbitrary vertical  
position by measuring films deposited with different diffuser positions.
We note that muon catalyzed fusion
reactions could also be used as a mono-energetic alpha source
for the thickness measurement~\cite{fujiw94} (see also
\cite{stras96}).

For the TRIUMF target system, the linearity of 
deposition was confirmed, with the
exception of very thin film deposition with system 2.
An asymmetric non-uniformity was observed in all films, while
deposition conditions (such as pumping of the high vacuum region
surrounding the target, 
or change in gas deposition rate) did 
not affect the thickness or uniformity.
Deposition from one side of the diffuser did not contaminate the cold
plate on the other side to a limit of one part in a thousand,
enabling experiments using two separated solid hydrogen 
targets~\cite{marsh93}. 
No evidence was seen for the change in the film thickness over time.
The comparison with Monte Carlo simulations 
indicates that the angular distribution of gas emission
is close to a $\cos \theta$ model.
The thickness calibration as well as film profile for 
the TRIUMF target is presented (Tables 1--3).

\section*{Acknowledgement}

The authors thank Professors D.F.~Measday, W.N.~Hardy and M.~Senba for
their valuable suggestions and discussions.
We gratefully acknowledge support of the Natural Sciences
and Engineering Research Council of Canada,
and the Swiss National Science Foundation.
M.C.F.\ thanks Rotary Foundation, the University of British Columbia
and the Department of Foreign Affairs and International Trade of the
Government of Canada for their support.

\newpage

\begin{table}[ht]
\begin{center}
\caption{Protium film thicknesses at various distance from the center of
the film for system 3. The distance between the diffuser surface and
cold plate surface was 8 mm.}
\label{tab:conversion}
\bigskip
\begin{tabular}{c|c} \hline
Position with respect to & Thickness per unit gas input \\ 
film center (mm) 
	  & ($\mu$g$\cdot$cm$^{-2}\cdot$(T$\cdot l$)$^{-1}$) \\ \hline
	40   & $0.21   \pm  0.08$ \\ 
	30   & $1.52   \pm  0.11$ \\
	20   & $2.92   \pm  0.17$ \\
	15   &  $3.10  \pm  0.18$ \\
	10   &  $3.21  \pm  0.16$ \\
	5    &  $3.33  \pm  0.19$ \\
	0    &  $3.46  \pm  0.17$ \\
	-5    &  $3.64  \pm  0.18$ \\    
	-10   &  $3.76  \pm  0.19$ \\     
	-15   &  $3.92 \pm  0.21$ \\
      	-20   &  $3.77 \pm  0.19$ \\
	-25   &	$3.06 \pm  0.17$ \\
	-30   &  $1.68 \pm  0.09$ \\ \hline
\end{tabular}
\end{center}
\end{table}

\newpage

\begin{table}[ht]
\begin{center} 
\caption{Thickness of the protium film for system 1, 
deposited with standard diffuser position.}
\label{tab:conversion_r1}
\bigskip
\begin{tabular}{c|c} \hline
Source position & Thickness per unit gas input \\
(mm) & ($\mu$g$\cdot$cm$^{-2}\cdot$(T$\cdot l$)$^{-1}$) \\ \hline
	20	&	$2.37 \pm 0.13$	\\
	10	&	$2.94 \pm 0.15$ \\
	0	&	$3.32 \pm 0.17$	\\ 
	-10	&	$3.60 \pm 0.19$ \\
	-20	&	$3.31 \pm 0.17$ \\ \hline 
\end{tabular}
\end{center}
\end{table}

\vspace*{2in}

\begin{table}[ht]
\begin{center} 
\caption{Thickness calibration factors for a neon film at source
positions -20 mm, 0 mm and 20 mm.}
\label{tab:neon}
\bigskip
\begin{tabular}{c|c} \hline
Source position & Thickness per unit gas input \\
(mm)		
  & ($\mu$g$\cdot$cm$^{-2}\cdot$(T$\cdot l$)$^{-1}$) \\ \hline
	20	&	$24.7 \pm 1.7$  \\
	0	&	$32.5 \pm 2.0$	\\ 
	-20	&	$31.5 \pm 1.9$  \\ \hline 
\end{tabular}
\end{center}
\end{table}

\newpage

\newpage

\newpage

\newpage

\newpage

\newpage

\newpage

\newpage

\newpage

\newpage

\end{document}